\documentclass[%
 reprint,
 amsmath,amssymb,
 aps,
]{revtex4-2}

\usepackage{graphicx}
\usepackage{dcolumn}
\usepackage{bm}
\usepackage{hyperref}
\usepackage{booktabs}
\usepackage{gensymb}
\usepackage[utf8x]{inputenc}                          
\usepackage[english]{babel}
\usepackage{float}
\usepackage{placeins}
\usepackage{subcaption}
\usepackage{orcidlink}
\usepackage{soul}   
\usepackage{xcolor} 
\usepackage{ragged2e}
\DeclareCaptionFormat{revjustify}{
  \justifying #1#2#3
}

\begin{document}

\preprint{APS/123-QED}

\title{Dipole Localization Using An Integrated Radio Frequency Atomic Magnetometer}

\author{Ayse Marasli\,\orcidlink{0000-0003-4449-0243}}

 \email{amarasli@gmu.edu}
\author{Karen L. Sauer\,\orcidlink{0000-0003-2784-8530}}%
 \email{ksauer1@gmu.edu}
\affiliation{ Quantum Science and Engineering Center, George Mason University, Fairfax, Virginia 22030, USA}
\author{Thomas Kornack\,\orcidlink{0000-0003-1615-5325}}
\email{kornack@twinleaf.com}
\author{Casey Oware\orcidlink{0009-0002-2681-3096}}
 \email{oware@twinleaf.com}
\affiliation{Twinleaf LLC, 300 Deer Creek Drive Suite 300, Plainsboro, NJ 08536, USA}

\begin{abstract}
Optically-pumped atomic magnetometers have previously been used in arrays to reject interference from far away sources and enable the sensitive detection of local sources of radio frequency (RF) signals, useful, for instance, in the detection of low field NMR signals in an unshielded environment. We now demonstrate a complementary scheme in which four magnetometer measurements are used to locate in three dimensions a nearby radio frequency source. The methodology relies on the measurement of a radio frequency vector at two different positions and modeling the source as a magnetic dipole of known orientation. In contrast to coil detection, magnetometers have the advantage of measuring signals in a 2D plane, and do not inductively couple to their environment or each other, making them a strong candidate for localization of hidden RF sources. For this demonstration, we use only a single RF magnetometer to make four measurements of a synchronous and oriented dipole source, but it is to be expected that this could be replaced by four magnetometers working simultaneously. In addition, this work is greatly aided by the introduction of a fully integrated magnetometer, in which all optics, including lasers, are safely enclosed into a compact head with flexible wired connections. The portable, as well as safe, nature of the sensor make it quite
valuable for in the field work.
\end{abstract}

\maketitle


\section{\label{sec:int}Introduction\protect }
Magnetometers find numerous applications~\cite{bai2023atomic}, ranging from measurement of DC, or close to DC fields, including archaeological magnetic prospecting~\cite{fassbinder2015seeing}, magneto-cardiography~\cite{bang2016repolarization, alem2015fetal} magneto-encephalography~\cite{hamalainen1993magnetoencephalography, alem2014magnetoencephalography, boto2018moving, nardelli2020conformal}
magnetic tracking~\cite{Xu2006, soheilian2020position, soheilian2021detection} and 
remanent detection of unexploded ordnance~\cite{billings2006magnetic, yoo2021application} (UXO); to radio-frequency measurements, including the search for axions~\cite{budker2014proposal},
induction detection of UXO \cite{zhang2003sensing},
low-field NMR~\cite{savukov2005nmr, savukov2007detection}, 
nuclear quadrupole resonance (NQR) detection of contraband material~\cite{garroway1994narcotics, Garroway2001, jenkinson2004nuclear, miller2005explosives, lee2006subfemtotesla, miller2011nuclear, cardona2015nuclear, Malone2025, ROBERTSON2025},
RF magnetic communication~\cite{gerginov2017prospects,fan2022magnetic}, and magnetic induction tomography (MIT) ~\cite{wickenbrock2014magnetic, Deans2018, deans2021electromagnetic}. Historically, detection of RF fields has been done through wire loops. Their sensitivity, however, degrades at lower frequency, and inductive and capacitive to their surroundings~\cite{suits2004noise} can corrupt the signal. Furthermore, it is common practice to boost the signal from the coil using a resonantly tuned circuit, whose reactive components must be mechanically changed when the frequency of interest changes.    Such mechanical changes inhibit search speeds when multiple frequencies are involved.
Atomic magnetometers (AMM), in contrast, retain their sensitivity at low frequency and can exceed that of a coil for frequencies below 50~MHz~\cite{savukov2007detection}.  Both this sensitivity, and their lack of inductive and capacitive coupling, make them attractive as measurement devices\cite{cooper2016atomic}. Additionally, their resonance frequency is easily tunable with a small static magnetic field~\cite{savukov2005tunable}, so that frequencies of interest can be rapidly detected \cite{quiroz2022interleaved} or even simultaneously measured with the appropriate geometry \cite{heilman2024large}. Moreover, these magnetometers have become more robust and flexible with integration of lasers and optics into a compact sensor head with only wired connections.  In this work, we present the use of an integrated RF magnetometer to localize an oriented magnetic dipole, the dipole representing the lowest order approximation to any magnetic source. 

In many of the magnetometer applications, the object of detection is out of the line of sight.  A common method of dipole localization relies on the measurement of the magnetic field and the magnetic field gradient tensor at a single point distant from the dipole; initial demonstration at 1~kHz was done with three loop coils and three gradiometer coils with a 500 turn solenoid for a field source~\cite{nara2006closed}, the number of turns reflective of the poor sensitivity of coil detection at such low frequency.  The chief advantage to this technique is that it is insensitive to the dipole orientation, the disadvantage is that the dipole must be far enough away that higher order terms are not significant.  Indeed later variations of this technique focus on obtaining higher order tensors for better localization, but at the expense of more computational and experimental complexity, for instance, the use of five tri-axial fluxgate magnetometers~\cite{gang2014magnetic}, or a fluxgate magnetometer mounted on a spinning disc that undergoes rotation about multiple axes\cite{sui2017multiple}.    Another variant measured the gradient tensors at two positions, shuttling a cross of four tri-axial fluxgate magnetometers between them, to get better positioning~\cite{xu2021magnetic}.  This demonstration, along with the others using fluxgate magnetometers, were DC measurements with a permanent magnet as the field source, presumably to create a significant magnetic field;  fields produced were in the $\mu$T range~\cite{xu2021magnetic}.  

Other methods avoid the distance approximation and adopt a more brute force approach to magnetic dipole localization.  For instance, the signals from an array of magnetometers can be directly fit to the magnetic dipole formula~\cite{Hashi2006}.  While positional accuracy from a 25 coil array operating at 306~kHz has shown to be quite good~\cite{hashi2011wireless}, the non-linear fitting of such data sampling would be time-consuming. Other researchers have taken on even larger arrays and more sophisticated search algorithms~\cite{de2022compressed}.  Of note, coil detection relies heavily on creating arrays in which mutual inductance is negligible;  some of the more complex field measurements, as have been done with the fluxgate magnetometers, would be difficult to duplicate with field coils operating at radio-frequencies because of this issue of mutual inductance. 

The measurement technique described here does not rely on the approximation of the fields at distant points, nor takes a brute force approach, rather it relies on the inversion of the dipole magnetic field to a direction vector at only two positions.  Therefore more complex setups are not needed to include higher order terms.  It does have the distinct disadvantage, however, of needing a fixed dipole direction, but there are applications in which this is not a hinderance, including those in which the dipole is created by excitation as part of the measurement technique, for instance, in NMR and NQR. The technique takes advantage of the sensitivity of the RF atomic magnetometer to signals in a 2D plane orthogonal to the tuning field aligned along the pump direction.  By combining two measurements, but with orthogonal pump directions, the RF magnetic field vector can be determined.  In principle two sensors could be used simultaneously to find the vector at a given location.  For our demonstration, however, we use only a single integrated magnetometer and make four measurements of the RF field from a timed AC dipole, with a pair of measurements made in two separate places.  From these atomic measurements, two direction vectors pointing at the nearby dipole are found and their projected intersection reveals the dipole's position. The frequency used for the demonstration was 423~kHz, corresponding to the NQR frequency of ammonium nitrate. 

\FloatBarrier
figure
\section{\label{sec:theory}Theory\protect }
\begin{figure}
\captionsetup{
  format=revjustify,
  singlelinecheck=false
}
\centering
\includegraphics[width=0.35\textwidth]{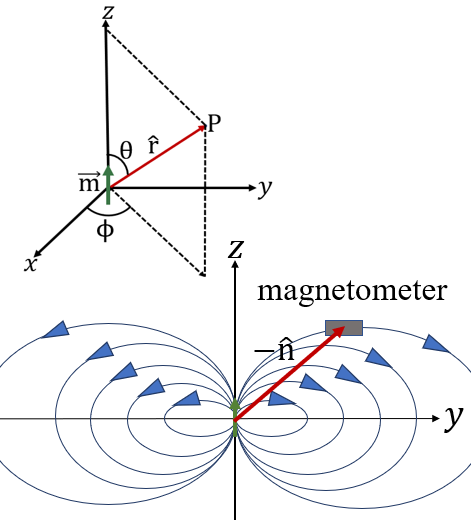}
\caption{\label{fig:magnetic_moment} Magnetic field lines from an idealized magnetic dipole.The angles $\theta$ and $\phi$ are defined as in the inset, while the unit direction vector $\hat{n}$ is from the magnetometer to the dipole.}

\end{figure}
The lowest order term in the multipole expansion of the magnetic vector potential of any
localized current distribution, including bound current sources, is that from a magnetic dipole Fig. \ref{fig:magnetic_moment}.  The magnetic field generated by an ideal magnetic dipole is 
\begin{equation}
\vec{B}_{\text{dip}}(\mathbf{r}) = \frac{\mu_0}{4 \pi r^3} \left[ 3(\vec{m} \cdot \hat{r}) \hat{r} - \vec{m} \right],
\label{dipole_moment}
\end{equation} 
where $\vec{m}$ denotes the magnetic dipole moment, 
$\vec{r}$ is the vector directed from the dipole to the measurement point $P$ and $\mu_0$ is the vacuum magnetic permeability. For consistency with the measurement configuration used in this study, the unit vector $-\hat{r}$ will hereafter be denoted as $\hat{n}$. When the orientation of the dipole is known and constrained to a half-space either above or below the dipole, as defined by the \textit{x/y}-plane in Fig. \ref{fig:magnetic_moment} , the direction of the magnetic field at $P$ can be used to determine $\hat{n}$ via Eq.~\ref{dipole_moment}.
Such spatial constraints are typical for dipoles induced by localized excitation, as is standard for example, in nuclear magnetic resonance (NMR) measurements. Assuming a dipole moment aligned along the $z$-axis, $\vec{m}=m\hat{z}$ , $\hat{r}$ in Eq. (1) can be rewritten in spherical coordinates as:
\begin{equation} 
\hat{r} = 
\begin{bmatrix}
\sin\theta \cos\phi, & \sin\theta \sin\phi, & \cos\theta
\end{bmatrix}.
\label{eq:vector_components}
\end{equation}
From $\cos^2\theta=\frac{1}{1+\tan^2\theta}$ relation,  $\vec{B}$ is written in terms of $\mathrm \theta$ and $\mathrm \phi$:

\begin{align}
\vec{B}_{\text{dip}}(r) &= \frac{3 m \mu_0}{4 \pi (1 + \tan^2 \theta) r^3} 
\begin{bmatrix}
\tan \theta \cos \phi \\
\tan \theta \sin \phi \\
\frac{2 - \tan^2 \theta}{3}
\end{bmatrix}
\end{align}

The ratio of transverse components gives $\mathrm \phi$:
\begin{equation} 
\phi={\mathrm {atan2}(B_{y},B_{x})},
\label{eq:phi}
\end{equation}                   
while the component $B_\perp = \sqrt{B_x^2 + B_y^2}$ and $B_z$ are used to find $\mathrm \theta$:
\begin{equation} 
\theta = \mathrm{atan2}{ \left[ - 3 B_z + \sqrt{9 B_z^2 + 8 B_\perp^2}, 
2 B_\perp \right]}.
\label{eq:theta_expression}
\end{equation}
Note, only the ratio of the fields, not their absolute values, are needed to determine the direction vector $\hat{n}$, and, as is discussed below in more detail, two spatially separated direction vector determine localization. Consequently, the calculation does not depend on the size of the magnetic dipole and the localization would be insensitive to attenuation by material barriers, for example soil for buried landmines \cite{Hibbs2001situ,cardona2015nuclear}.  
Therefore {to determine  $\hat{n}$, the relative ratio of all three components of the magnetic field vector must be measured. If the timing of the dipole is known, as would be the case with many techniques in which excitation drives a response, the three components can be found with two consecutive measurements by a single magnetometer, with a $90^\circ$ rotation done between measurements. In our experiment the first measurement was taken with the tuning field aligned along $z$, and the probe beam along $x$ allowing $B_x$ and $B_y$ to be determined. The projection of the RF field on to 2D plane orthogonal to $z$ can be expressed as
\begin{equation}
B_{RF\perp} = \frac{2\omega_{1\perp}}{\gamma}\cos\{\omega(t-t_0)\}(\hat{x}\cos\eta + \hat{y}\sin\eta)
\label{eq:B_RF}
\end{equation}
where $\gamma=2\pi(700\,\text{kHz/G})$ is the gyromagnetic ratio, and the unit vector combination $(\hat{x}\cos\eta+\hat{y}\sin\eta)$ specifies the transverse field orientation. Using a circularly polarized pump beam resonant with the atomic D1 line and aligned with the tuning field, a net atomic polarization is created along the tuning field. The RF fields tip some of the polarization into the transverse plane, which then precess around the tuning field. The magnetometer signal is directly proportional to the atomic polarization along the probe beams $P_x$:\cite{Alem2011}

\begin{equation}
P_x = \omega_{1\perp}T_2 \left[ \frac{1}{\sqrt{1+\Delta\omega^2 T_2^2}} \right]
\sin\{\omega(t-t_0) - \Delta\omega\ t + \eta\}
\label{eq:P_x}
\end{equation}
and phase-sensitive detection, enables the spectrometer to resolve the sinusoidal phase. In this expression, both the quantity in square brakets and the phase shift $\Delta\omega t$ depend on the off-resonance of the excitation frequency from the Larmor frequency
\begin{equation} 
\Delta\omega=\omega-\gamma B_0
\label{eq:Larmor}
\end{equation}
where $B_0$ is the static magnetic field at the sensor. Such resonance effects are corrected using a calibration signal, for by the parameter defining orientation, $\eta$, to be determined. Once $\eta$ is known, $B_x$ can be distinguished from $B_y$. The sensor is then rotated so that the pump beam lies along $x$, enabling measurement of $B_y$ and $B_z$. For arbitrary timing, simultaneous measurements with multiple sensors would be required to capture all three components.Furthermore,the technique exploits the RF atomic magnetometer’s sensitivity to signals in a 2D plane orthogonal to the tuning field. Transverse RF fields generate the signal by tipping the net magnetization of optically pumped atoms out of alignment with the static field.\par
Upon the vector directions, Eqs.~\ref{eq:vector_components}, \ref{eq:phi}, and \ref{eq:theta_expression}, from two separation positions are known, the dipole displacement $\vec{d}$ can be determined based on the intersection of lines extending along their direction vectors, $\hat{n}_1$ and $\hat{n}_2$.
Figure~\ref{fig:Dipole_Position} represents the vectorial schematic of the dipole, with respect to the two positions of measurement $\mathrm S_1$ and  $\mathrm S_2$. The displacement between the two sensors is denoted as $\bigtriangleup\hat{x}$, and the displacements between the sensor positions and the dipole are defined as $\vec{V_1}$, and  $\vec{V_2}$, respectively.
\begin{figure}[htbp]
\captionsetup{
  format=revjustify,
  singlelinecheck=false
}
\includegraphics[width=0.35\textwidth]{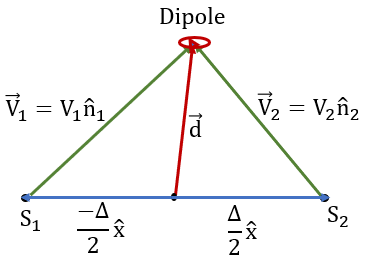}
\caption{\label{fig:Dipole_Position} Vectorial schematic of the dipole position with respect to  $\mathrm S_{1}$ and $\mathrm S_{2}$ representing the positions of the alkali metal cell of each sensor and the origin is taken on the \textit{x}-axis at a position equidistant from $\mathrm S_{1}$ and $\mathrm S_{2}$. The \textit{x}-axis is uniquely defined as the direction between the two sensors.}
\end{figure}

As can be seen in Fig.~\ref{fig:Dipole_Position}, $\vec{d}$ can be written as:
\begin{equation}
	\vec{d} = \frac{\vec{V}_1 + \vec{V}_2}{2}= \frac{V_1 \hat{n}_1 + V_2 \hat{n}_2}{2},
	\label{position}
\end{equation}
and 
\begin{equation}
	\bigtriangleup\hat{x}= \vec{V}_1 - \vec{V}_2 .
	\label{deltax}
\end{equation}
Therefore, $\bigtriangleup$ can be expressed in terms of the magnitudes of $V_{1}$ and  $V_{2}$ :
\begin{subequations}
	\begin{align}
		\bigtriangleup = n_{1x}V_{1}-n_{2x}V_{2},
        \label{eq:8a} 
	\end{align}
where the unit vectors $\hat{n}=[n_x,n_y,n_z]$ as shown in  Fig. \ref{fig:magnetic_moment}. Further, from inspection of Fig. \ref{fig:Dipole_Position}, it is clear that the components of $\vec{V}$ orthogonal to $\hat{x}$ are equal,
\begin{align}
\hat{n}_{1\perp} V_1 = \hat{n}_{2\perp} V_2,
\label{eq:8b}\\
|n_{1\perp}| \, V_1 = |n_{2\perp}| \, V_2.
\label{eq:8d}
\end{align}	
\end{subequations}
where $\hat{n}_{1\perp}=[0,n_{1y},n_{1z}]$ and $\hat{n}_{2\perp}=[0,n_{2y},n_{2z}]$. \newline From combination of \ref{eq:8a}-\ref{eq:8d}, $\bigtriangleup$ can be re-expressed as:
\begin{equation}
	\Delta = \left(n_{1x} \frac{ |n_{2\perp}|}{|n_{1\perp}|} - n_{2x} \right) V_2 
	= \left( n_{1x} - n_{2x} \frac{|n_{1\perp}|}{|n_{2\perp}|} \right) V_1.
\end{equation}
Consequently, using Eq.~\ref{position},
$\vec{d}$ can be written solely in terms of $\bigtriangleup$, and the two directional unit vectors,
\begin{equation}
	\vec{d} = \frac{\Delta}{2} \left[ \frac{\hat{n}_1 |n_{2\perp}|+\hat{n}_2 |n_{1\perp}|}{n_{1x} |n_{2\perp}| - n_{2x} |n_{1\perp}|}\right].
\end{equation}
\FloatBarrier
\section{\label{sec:setup}Experimental Setup\protect }

\subsection{\label{sec:level2}Characterization of Integrated RF Atomic Magnetometer\protect}
At the core of the integrated sensor head is a cell measuring $3.8\times3.8\times9$~mm$^3$, containing $^{87}$Rb atoms, 0.7 amagat of neon as a buffer gas and 0.1 amagat of the quenching gas N${_2}$.  A resistive heating oven creates a saturated vapor pressure, with close to $4 \times 10^{13}$~atoms/cm$^3$.  The optical system employs a crossed pump–probe configuration  \cite{RevModPhys.44.169,budker2007optical}. The pump laser is tuned to the D1 transition, while the probe laser is de-tuned by 0.16 nm below the D1 line to maximize the signal size.  Both beams are collimated, with a quarter wave plate producing circular polarization of the pump and a linear polarizer ensuring linear polarization of the probe.  The pump beam traverses the vapor cell once, whereas the probe beam is reflected by a mirror to achieve a double pass  \cite{cooper2016atomic}, thereby enhancing the polarization rotation angle \cite{PhysRevLett.18.577}.  The resulting rotation, proportional to the resonant RF field \cite{savukov2005nmr}, was measured with a balanced polarimeter and photo-diode detection\cite{chauvat1997magnification,makarov2023observation}. Surrounding these elements are tri-axial coils used to define the bias field and set the resonance frequency, with all components housed in a tube measuring 15 cm in length and 4 cm in diameter, as illustrated in Fig.~\ref{fig:sensor}(a). The complete sensor system consists of the sensor head, differential amplifier for the polarization signal, and a bench-top current supply for controlling the current in the three coils with LabVIEW. Input and output signals are transmitted using electrical cables instead of fiber optic cables, which is an important detail for ease of use and robustness Fig.~\ref{fig:sensor}(b).

\begin{figure}[htbp]
\captionsetup{
  format=revjustify,
  singlelinecheck=false
}
\centering
\subfloat[]{\includegraphics[width=0.3\textwidth]{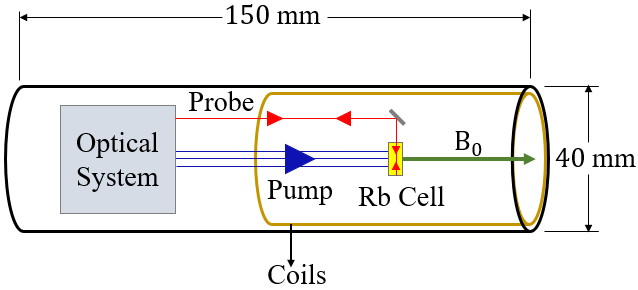}}%
\hfill
\subfloat[]{\includegraphics[width=0.2\textwidth]{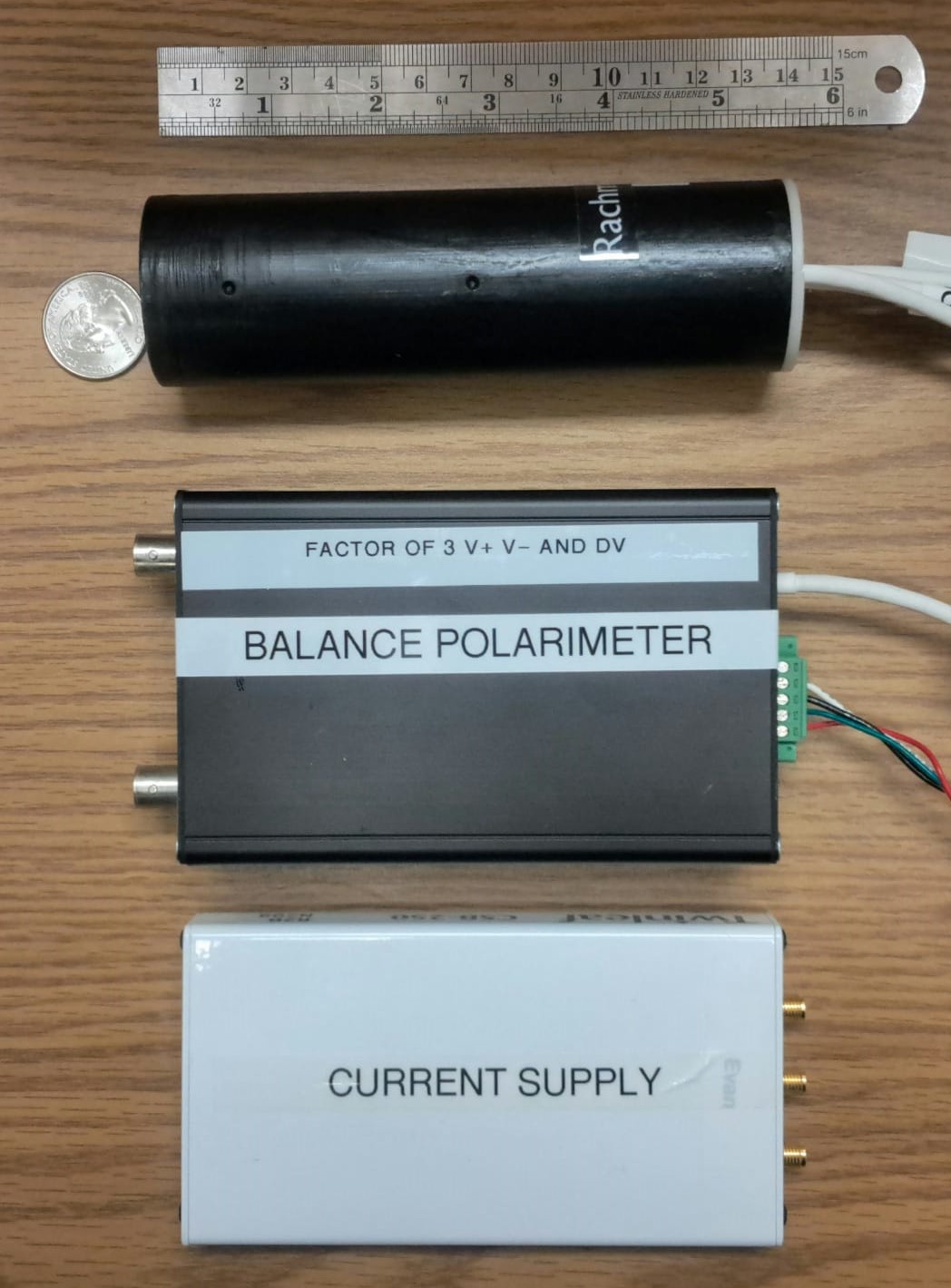}}
\caption{\label{fig:sensor}a) This schematic shows a top view of the sensor, with an atomic cell (yellow), crossed pump (blue) and probe (red) laser beams, and field coils (brown) creating the bias field (green). b) All these components are enclosed in a cylindrical black tube. Also shown is a bench-top current controller, and the differential amplifier for use with the balanced polarimeter.}
\end{figure}

\begin{figure}[htbp]
\captionsetup{
  format=revjustify,
  singlelinecheck=false
}
\centering
\subfloat[]{\includegraphics[width=0.45\textwidth]{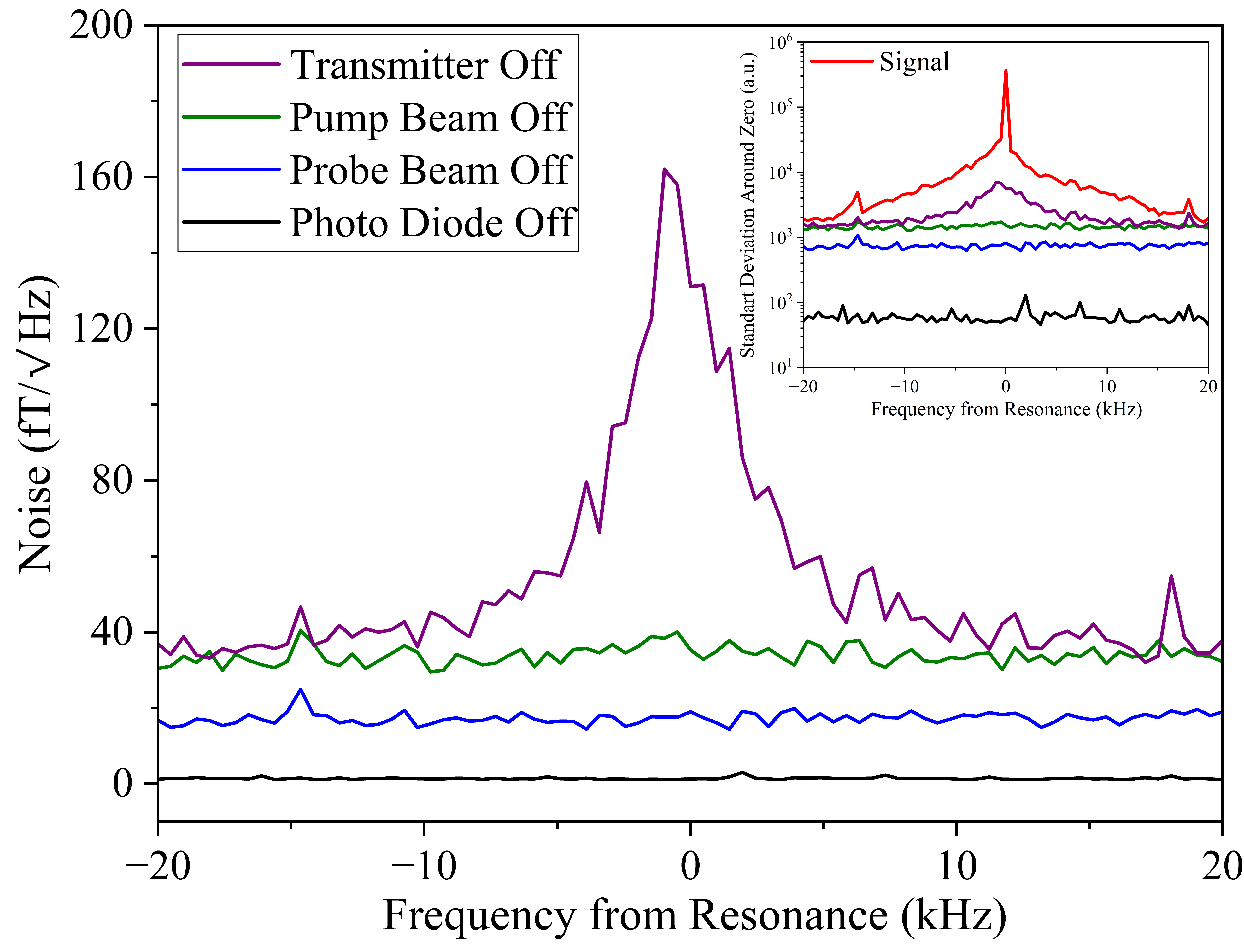}}
\hfill
\subfloat[]{\includegraphics[width=0.45\textwidth]{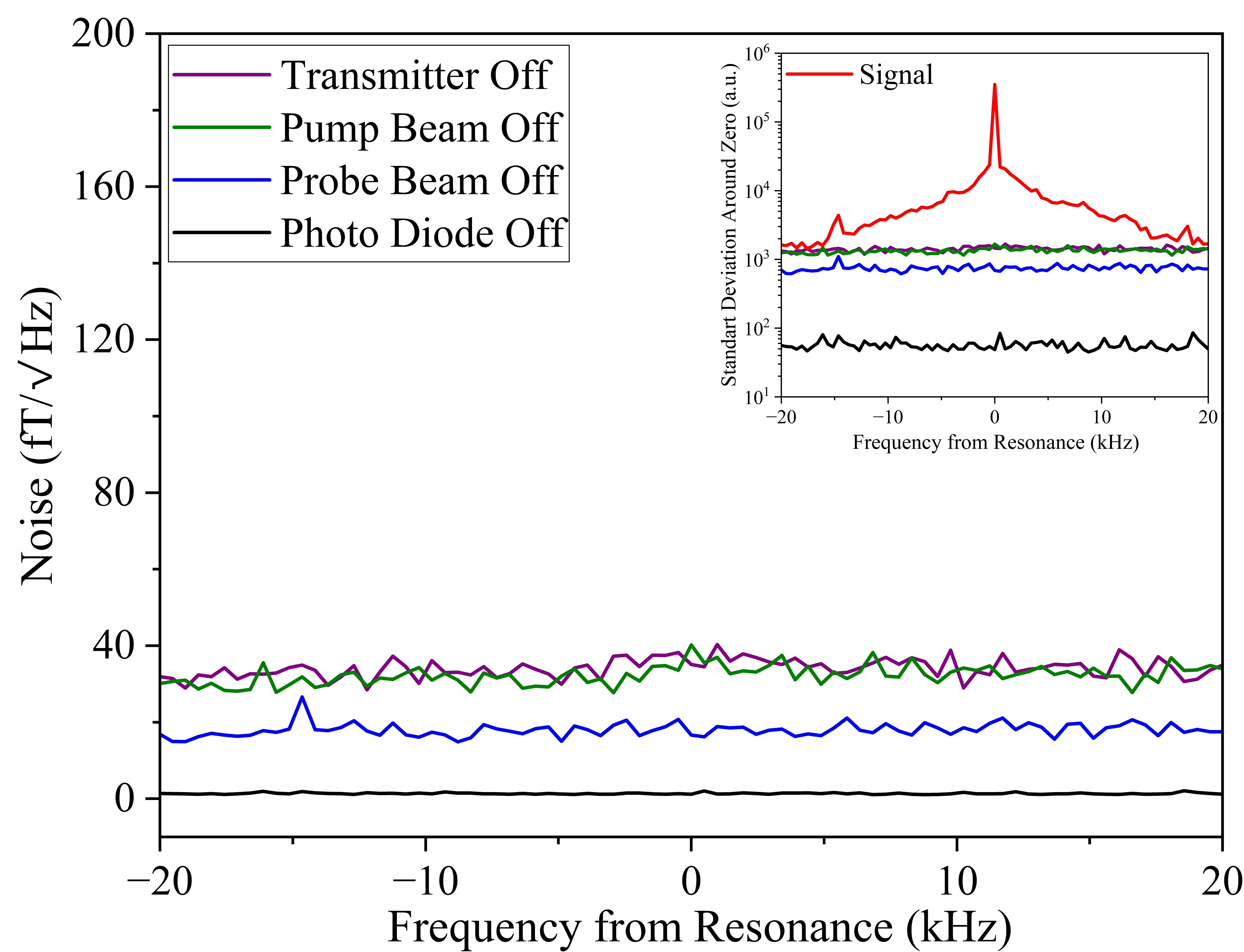}}
\caption{\label{fig:sensitivity} Noise spectra measured at the resonance frequency of 423 kHz (NQR of ammonium nitrate) with an acquisition time of 2.048 ms.   Signals obtained with an active RF transmitter driving the magnetometer with 0.2 nT resonant excitation are shown in red (inset).   Spectrometer noise (black), with additional electronic noise (blue), and further added probe beam noise (green) exhibit similar behavior both (b) with and (a) without shielding. Magnetic noise or sensitivity (purple), observed with optically pumped atoms, is markedly higher without shielding due to environmental laboratory noise.}
 \end{figure}

After optimization of the signal with respect to the laser parameters, the spin-spin relaxation time $T_{2}$ of a free induction decay was measured to be 0.17~ms. Moreover, as illustrated in Fig. \ref{fig:sensitivity}, the contributions of different noise sources were identified by comparing noise spectra under varying conditions. A small amount of white noise is present when only the spectrometer is active. Activating the electronics of the balanced polarimeter introduces an additional white noise component. When the probe beam is switched on, photon shot noise becomes evident. Finally, with the pump engaged and the atoms optically pumped, magnetic noise from the surrounding environment is observed at the resonant frequency. In the absence of RF shielding, a strong ambient magnetic signal dominates at the resonant frequency; however, placing the sensor inside a large aluminum enclosure suppresses this interference. Under these conditions, the noise is primarily dominated by photon shot noise, yielding a sensitivity of $35\pm 3\quad\mathrm{fT}/\sqrt{\mathrm{Hz}}$.

\subsection{Magnetic Dipole Measurement Setup\protect}
As shown in Fig.~\ref{fig:setup}(a), two Garolite sheets were arranged at right angles to each other, with the bottom plane dedicated to magnetometer measurements.  A five-turn coil with a diameter of 2~cm was positioned $L = 19.5$ cm above this sensor plane to mimic a covert source of signal, for instance a landmine in the ground detected through NQR~\cite{Garroway2001, Barrall2004}.  The two measurements positions, were spaced $2 L$ apart, and sat at the corners of the bottom plate.  Sensor platforms were constructed to permit precise alignment of the sensor through pitch, yaw, and roll adjustments. In addition, a three-turn coil was
wound along the edge of the bisecting plate to create a
continual source of calibration across both sensor positions to mitigate the effects of a changing ambient magnetic field in an unshielded environment; excitation and data acquisition were alternated between the dipole and the calibration coils.
This type of universal calibration would also be integral to measurement integrity if multiple sensors were used instead of multiple measurements \cite{cooper2016atomic,cooper2018rf}; the calibration would automatically compensate for the effects of stray residual tuning fields from other sensors. 
\begin{figure}[htbp]
\captionsetup{
  format=revjustify,
  singlelinecheck=false
}
\centering
\subfloat[]{\includegraphics[width=0.35\textwidth]{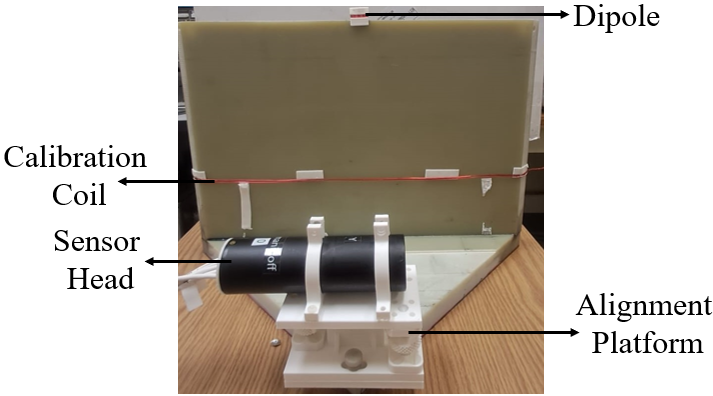}}%
\hfill
\subfloat[]{\includegraphics[width=0.35\textwidth]{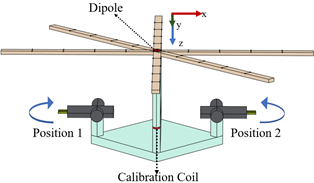}}
\hfill
\subfloat[]{\includegraphics[width=0.35\textwidth]{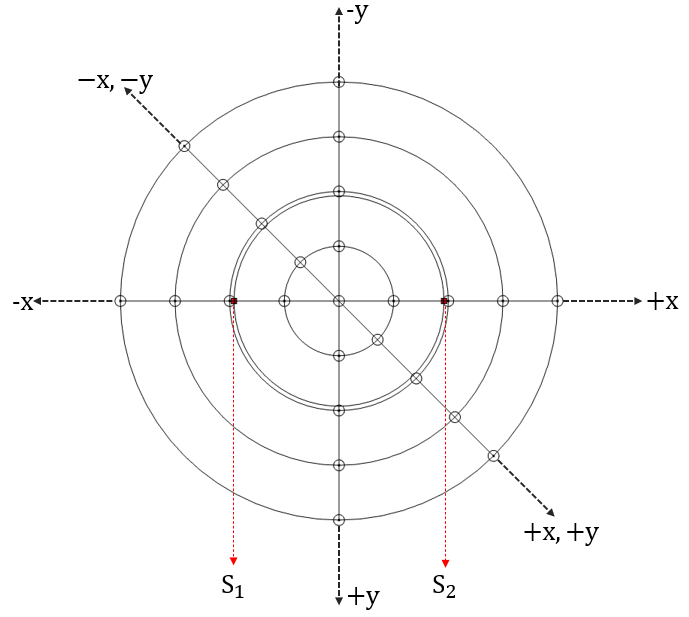}}
\caption{\label{fig:setup} a) Photo of the sensor head, alignment platform, dipole and calibration coils. b) A small coil with AC current running through it at 423~kHz is moved along a ruler above the plane containing the magnetometer positions. Measurement was taken in the positions shown, as well as with the sensor rotated by 90 degrees. c) Top view of measurement points and sensor positions during dipole movements along (i) only the \textit{x}-axis, (ii) only the \textit{y}-axis and (ii) the diagonal \textit{x/y}-axis.}
\end{figure}

Two separate measurements for each location were taken from the left side (position 1) and right side (position 2) corners of the structure shown in Fig.~\ref{fig:setup}(b), with the probe direction of the sensor perpendicular and parallel to the calibration coil. Both excitation of the RF coils and acquiring data were done with a phase-sensitive spectrometer from Tecmag~\cite{TECMAG}. Experimental repetition was synchronized with AC line to avoid unnecessary phase jitter. According to laboratory coordinate system, three different data sets were obtained by orienting the ruler along the \textit{x}-axis, \textit{y}-axis and \textit{x/y}-axis as seen in Fig.~\ref{fig:setup} (b-c).
\FloatBarrier
\clearpage
\section{\label{sec:result}Result and Discussion\protect }

\subsection{Positioning through magnetic field measurements\protect}

The measured and predicted magnetic field components ($B_{x}$,$B_{y}$,$B_{z}$) of movement of the magnetic dipole along the \textit{x}-axis,\textit{y}-axis and \textit{x/y}-axis at two sensor positions, $\mathrm{S_{1}}$ (red) and $\mathrm{S_{2}}$ (blue), are shown in Fig. \ref{fig:B_field} from (a) to (i), respectively.  As the dipole moves along the \textit{x}-axis and passes directly over a sensor’s position, a strong $B_z$ magnetic field is measured in that sensor as shown in Fig.~\ref{fig:B_field} (a-c). Close to sensor, but on either side of it, a strong $B_x$ field is  also measured flipping from one sign to another as expected. Reduced $\chi^2$ values for the $B_x$ and $B_z$ components are close to 1, reflecting strong agreement between theoretical and experimental data. However, the $B_y$ component, expected to be zero for all positions, shows deviations caused by dipole misalignment, on the order of a couple of degrees; careful inspection shows a faint pattern of $B_z$ reflected in the $B_y$ data.

As the dipole moves symmetrically between the sensors, along the \textit{y}-axis, the measured $B_y$ and $B_z$ fields are well matched and the $B_x$ components are opposite in sign as depicted in Fig.~\ref{fig:B_field}(d-f).  Reduced $\chi^2$ values range from 1 to 3, suggesting reasonable agreement between theoretical predictions and experimental results.
\begin{figure}
\captionsetup{
  format=revjustify,
  singlelinecheck=false
}
\centering
\subfloat{\includegraphics[width=0.45\textwidth]{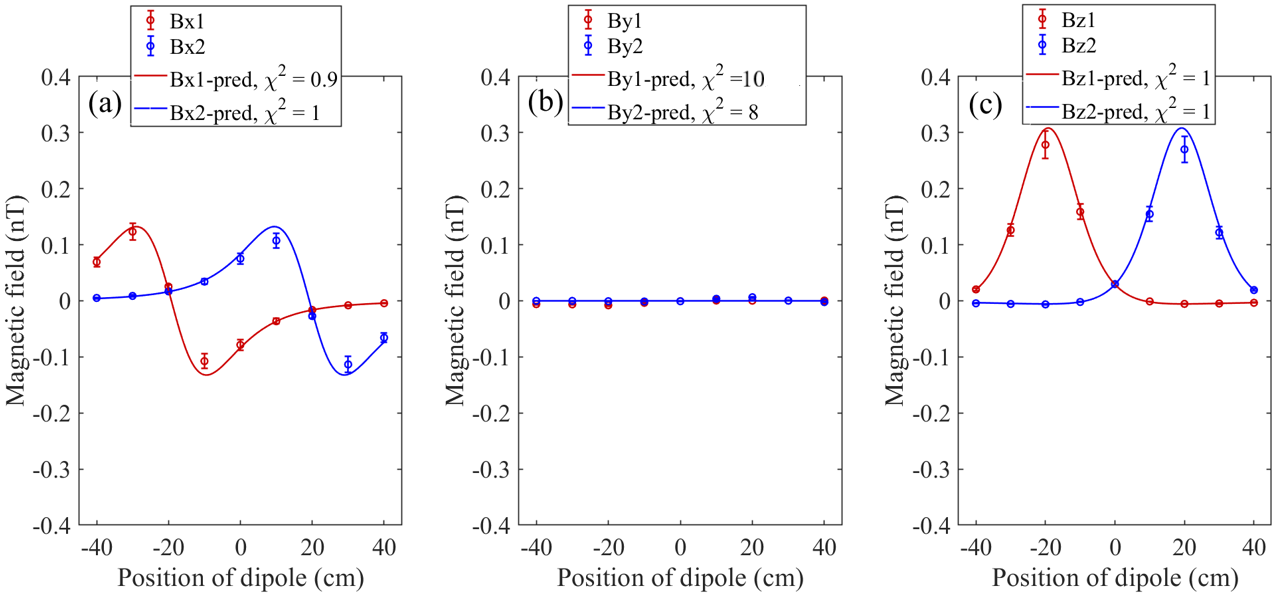}}

\subfloat{\includegraphics[width=0.45\textwidth]{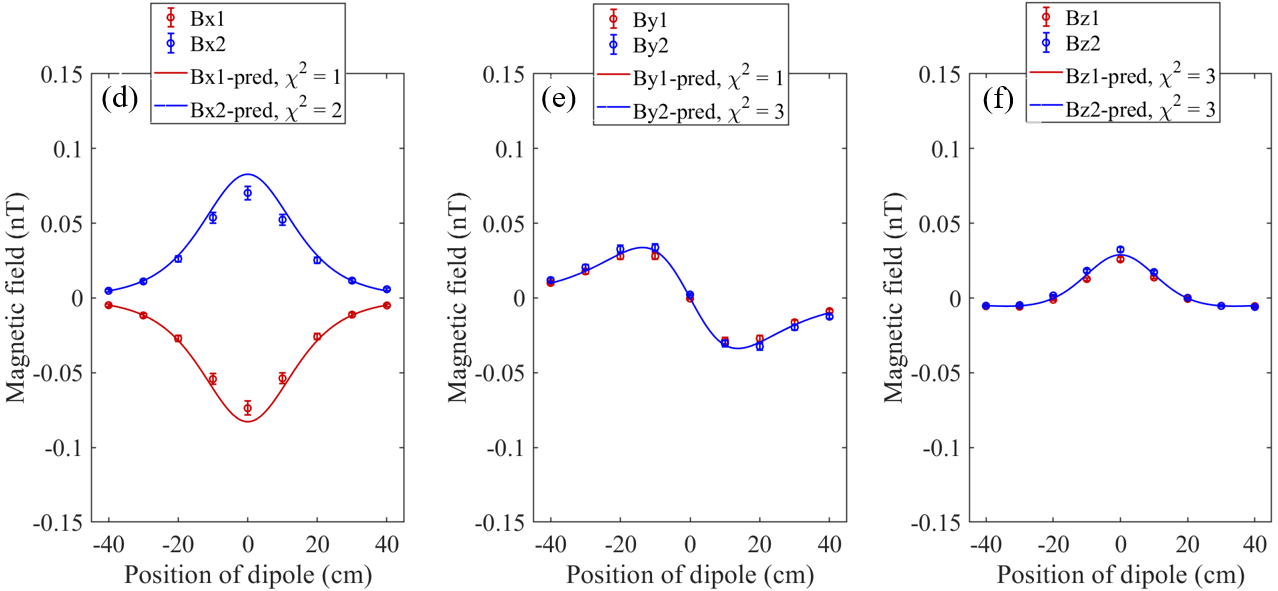}}

\subfloat{\includegraphics[width=0.45\textwidth]{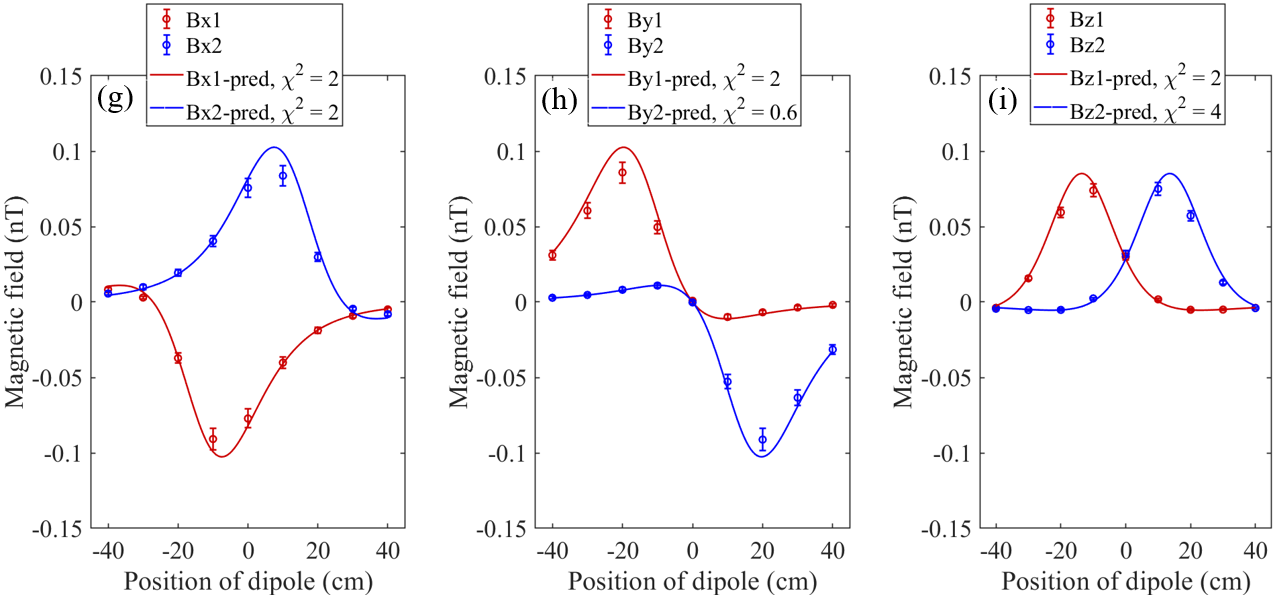}}
\caption{\label{fig:B_field}Comparison of the predicted and measured magnetic field components during dipole motion along   (a-c) \textit{x}-axis (d-f)  \textit{y}-axis, and (g-i)  \textit{x/y}-axis, respectively.The solid lines represent theoretical predictions obtained without the use of free parameters, while the experimental measurements are indicated by circular markers.}
\end{figure}

In the third experiment, the dipole moves diagonally with respect to the sensor positions. 
It moves from the upper left half quadrant of the Fig.~\ref{fig:setup}(c), closest to sensor position 1, to the lower right quadrant, closest to sensor position 2. The proximity effect can be observed in the peak position of Fig.~\ref{fig:B_field}(g-i).  The diagonal movement leads to having significant field components in all three directions as well as the observed change in field signs. Reasonable agreement is also observed in this data, but the prediction tends to overestimate field values suggesting a small scaling factor of the calibration might be appropriate. The localization methodology, however, does not depend on absolute value of the fields, rather only the field values relative to each other.

The angular variations of the magnetic field components, crucial for dipole localization, were systematically investigated for dipole motion along the \textit{x}-axis, \textit{y}-axis, and \textit{x/y}-axis, as presented in Fig.~\ref{fig:Angle}(a-f), respectively. The symbol $\phi$  represents the azimuthal angle of the magnetic field, calculated as $\phi = \text{atan2}(B_y, B_x) $, while the symbol $\vartheta$ represents the polar angle, calculated as $\vartheta = \text{atan2}(B_{\perp}, B_z)$ for both sensor positions, $\mathrm{S_1}$ and $\mathrm{S_2}$, in comparison with theoretical predictions.

\begin{figure}[htbp]
\captionsetup{
  format=revjustify,
  singlelinecheck=false
}
\centering
\subfloat{\includegraphics[width=0.45\textwidth]
{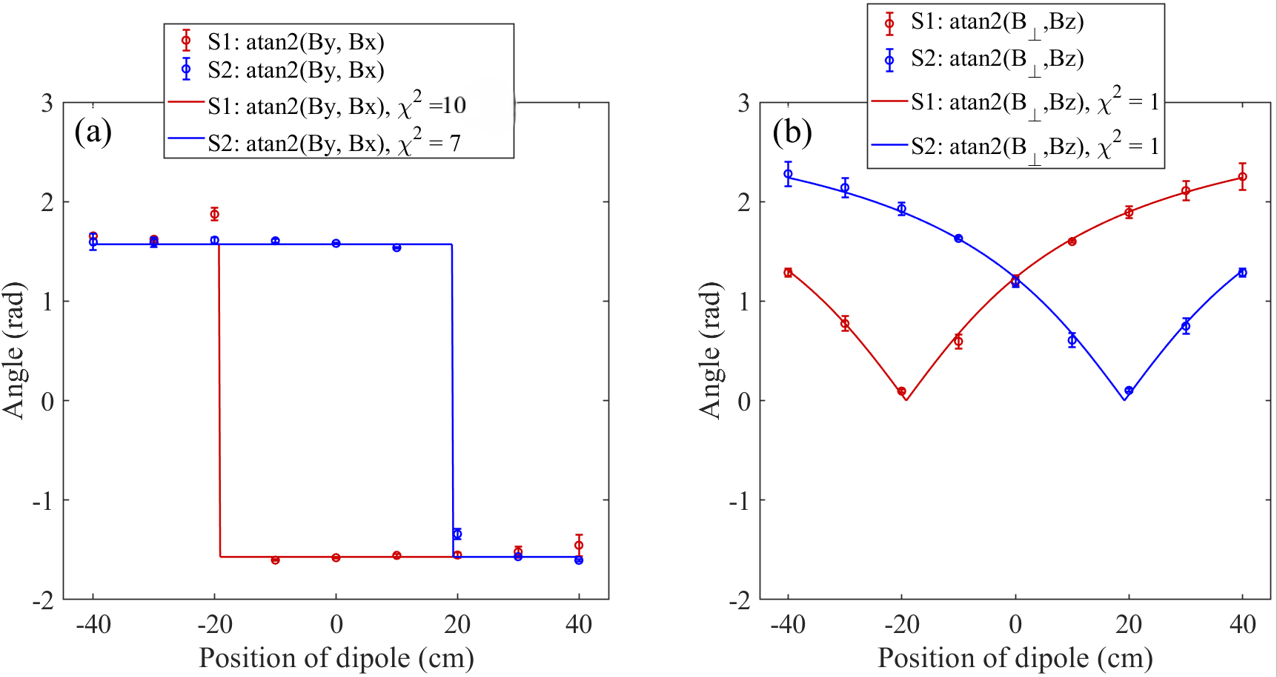}}

\subfloat{\includegraphics[width=0.45\textwidth]
{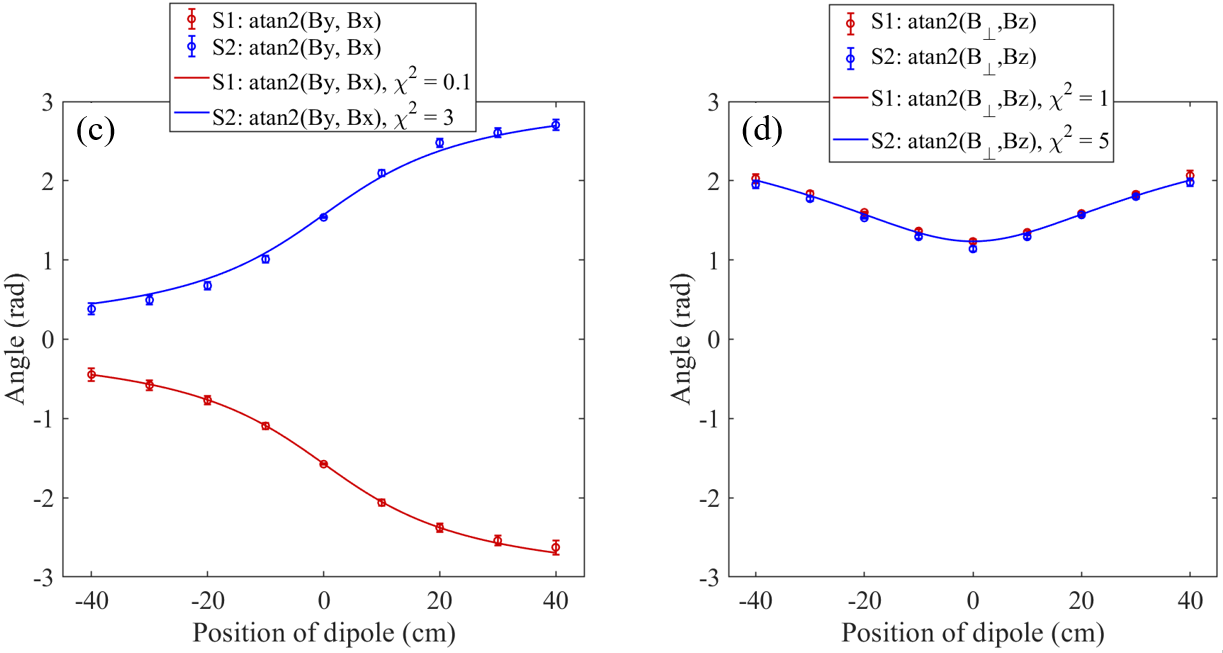}}

\subfloat{\includegraphics[width=0.45\textwidth]
{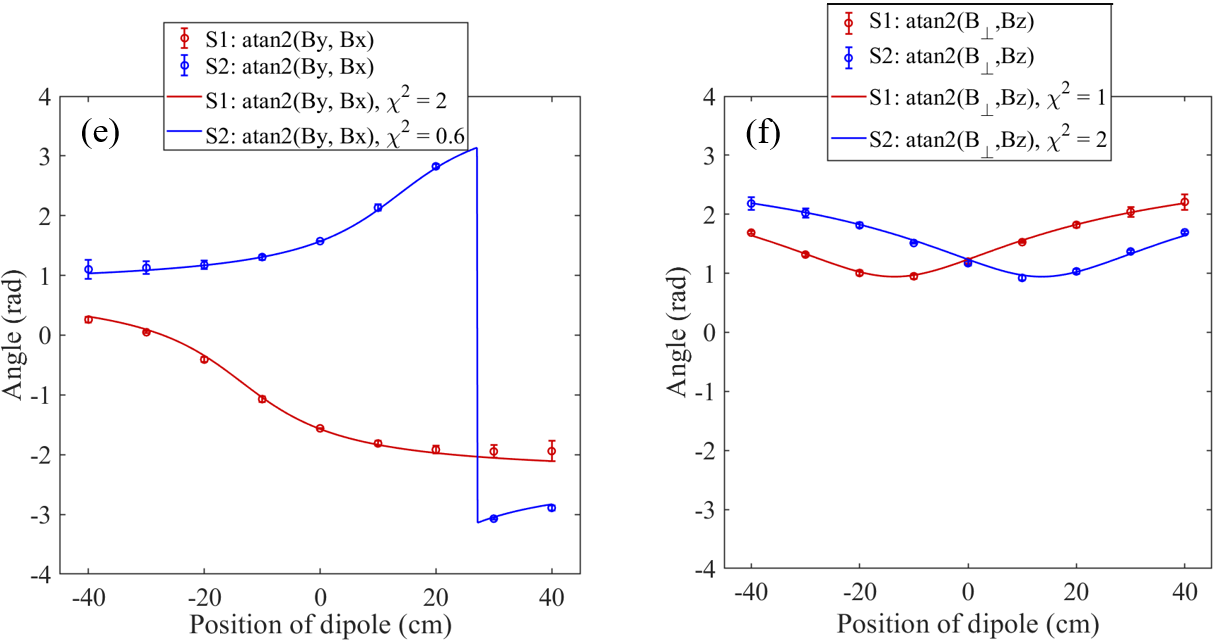}}

\caption{\label{fig:Angle}Predicted and measured angles between the magnetic field components for the motion of the dipole: (a) for $\phi$ along the $x$-direction, (b) for $\vartheta$ along the \textit{x}-direction, (c) for $\phi$ along the \textit{y}-direction, (d) for $\vartheta$ along the \textit{y}-direction, (e) for $\phi$ along the $x/y$-direction, (f) for $\vartheta$ along the \textit{x/y}-direction. The solid lines represent theoretical predictions obtained without the use of free parameters, while the experimental measurements are indicated by circular markers.}
\end{figure}

Overall, the experimental data show strong agreement with theoretical models, with reduced chi-square values close to one in most cases. Observed discrepancies are primarily attributed to environmental interference, dipole misalignment and idealized assumptions about the dipole. While these factors contribute to deviations, the general agreement remains robust. These results are used find the position of the dipole.
\begin{figure}
\captionsetup{
  format=revjustify,
  singlelinecheck=false
}
\centering
\subfloat[]{\includegraphics[width=0.4\textwidth]{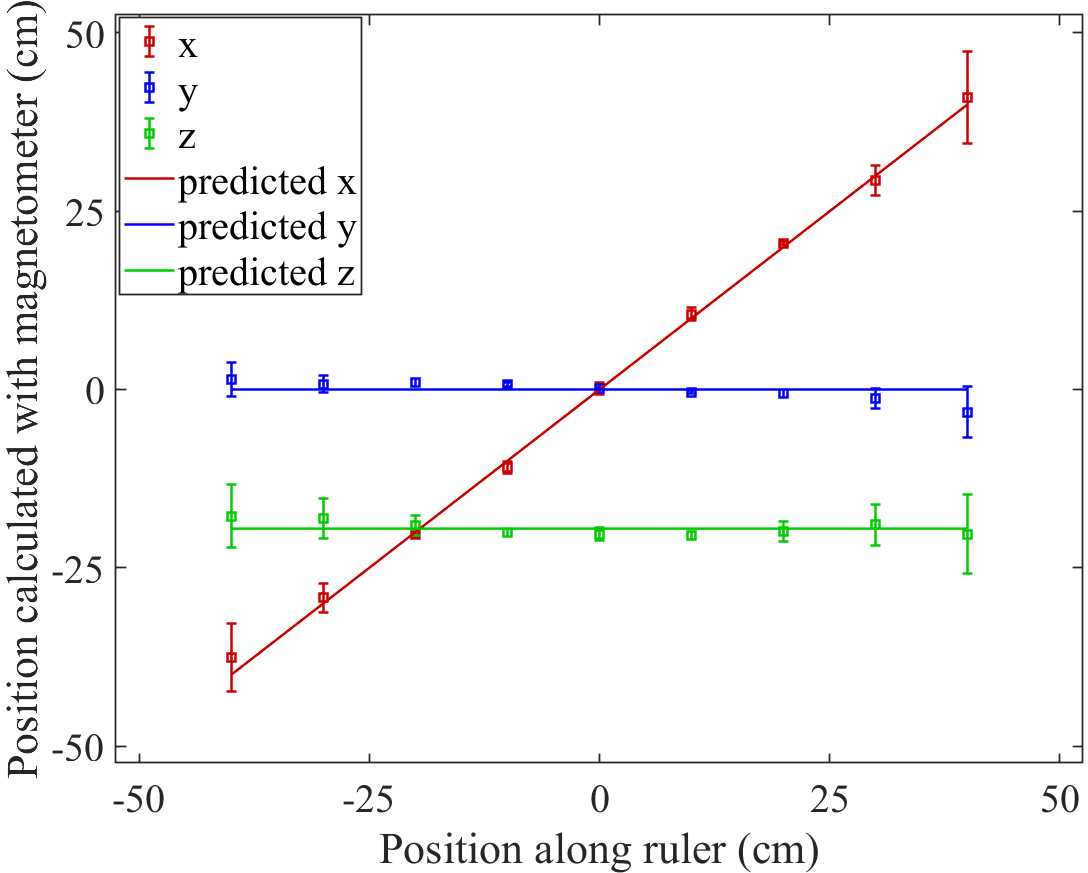}}

\subfloat[]{\includegraphics[width=0.4\textwidth]{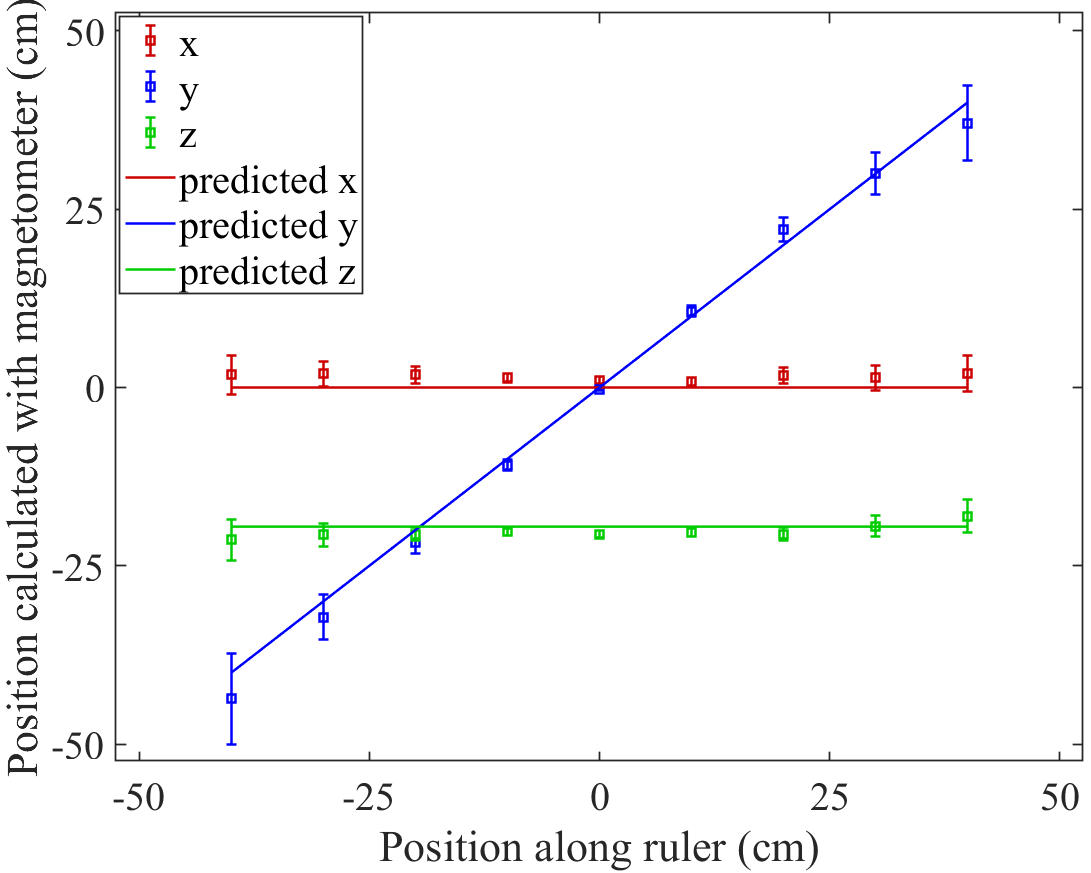}}

\subfloat[]{\includegraphics[width=0.4\textwidth]{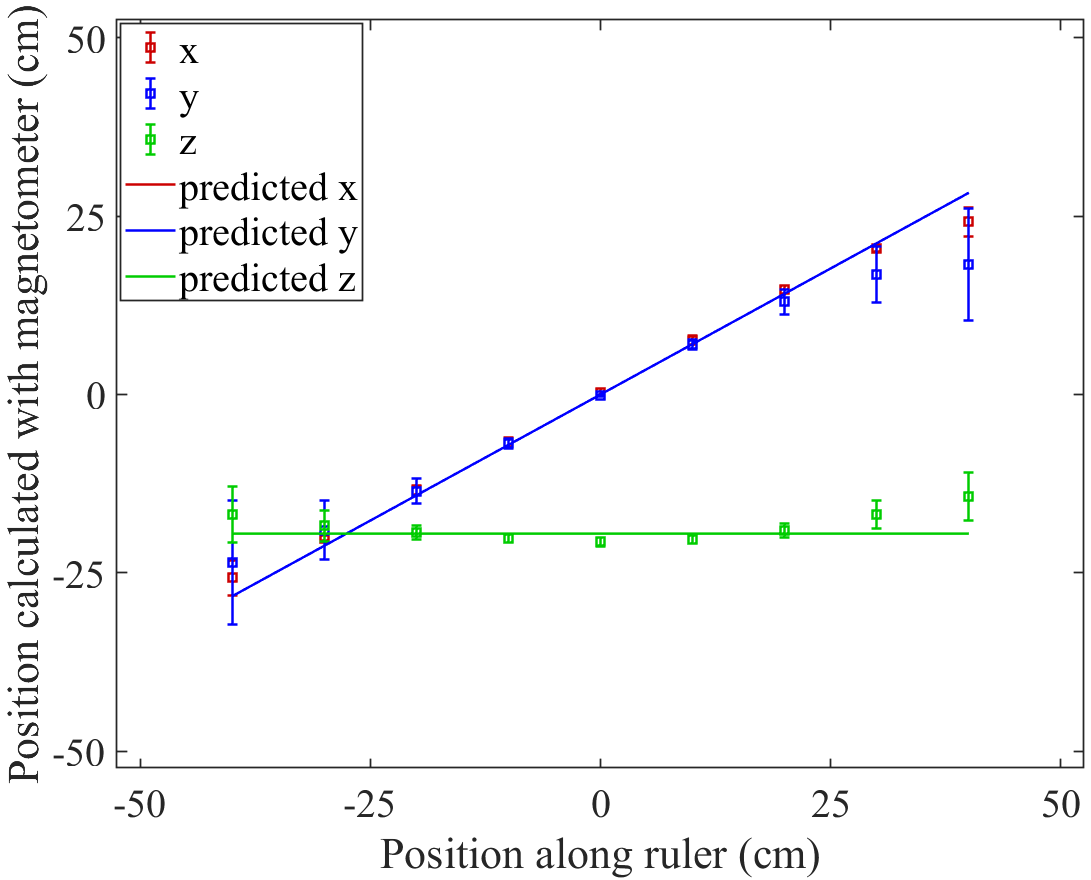}}
\caption{\label{fig:Position}Dipole coordinates $x$ (red), $y$ (blue), and $z$ (green) with respect to the midpoint between sensor positions with: (a) dipole motion along the \textit{x}-axis, (b) dipole motion along the \textit{y}-axis, and (c) dipole motion along the \textit{x/y}-axis. Solid lines indicate the actual dipole positions, while square markers represent the positions calculated from experimental data.}
\end{figure}

Figure~\ref{fig:Position} (a), (b), and (c) illustrate the coordinates of the dipole as calculated from magnetic fields alone as compared to the physical coordinates, for displacement exclusively along the \textit{x}-axis, the \textit{y}-axis, and along the diagonal, respectively.  Coordinates are with respect to the midpoint between the two measurement positions, and larger error bars reflect measurements made when the magnetic fields are weaker.  Overall there is good agreement between the position determined by the magnetometer and the physical position in all three experiments.   The motion along the ruler, is observed in the corresponding coordinate, that is $x$ in Fig.~\ref{fig:Position}(a) with $y$ at zero,  $y$ in Fig.~\ref{fig:Position}(b) with $x$ zero, and in both $x$ and $y$ in Fig.~\ref{fig:Position}(c).  In each, the constant displacement of the dipole from the sensor plane by $L = 19.5$~cm is observed through the $z$-coordinate.  To underline the potential use of the technique to locate a covert target in a three dimensional space, Fig.~\ref{3D_position}a plots the potential volume in which the dipole may be located according to the positions and error bars of Fig.~\ref{fig:Position}.Unsurprisingly, error in localization increases when the dipole is far away from the sensor, as the field strength from a magnetic dipole falls off as $1/r^3$, as described in Eq.\ref{dipole_moment}. Figure \ref{3D_position} b shows the localization error, taken from the mean of the error bars in Fig.\ref{fig:Position} at each position, scales roughly as $d^3$, where $d$ is the dipole’s distance from the origin, the midpoint between sensor measurement positions, as shown in Fig.\ref{fig:Dipole_Position}.
\begin{figure}[H]
\captionsetup{
  format=revjustify,
  singlelinecheck=false
}
\centering
\subfloat[]{\includegraphics[width=0.5\textwidth]{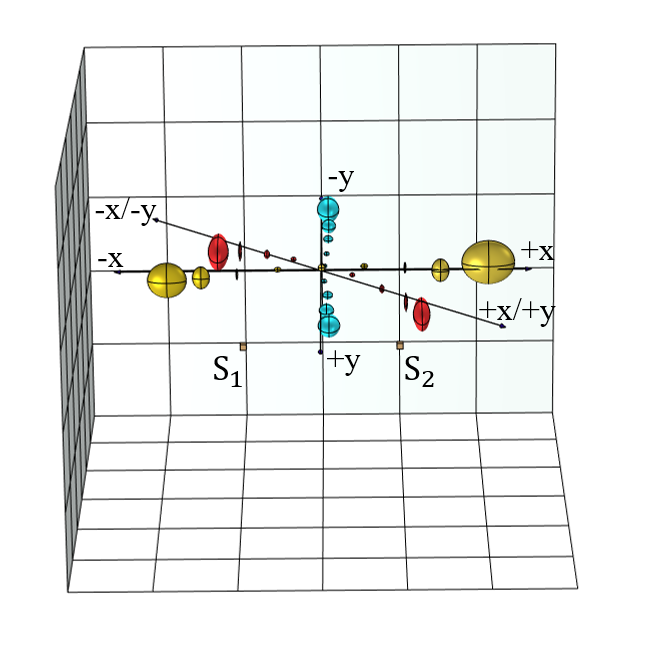}}

\subfloat[]{\includegraphics[width=0.45\textwidth]{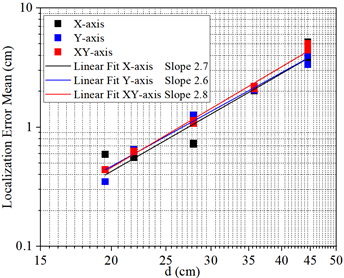}}
\caption{\label{3D_position} a) Three-dimensional representation of the potential volumes where the dipole may be located. Yellow indicates possible dipole positions along the \textit{x}-axis, red represents positions along the \textit{y}-axis, and blue corresponds to positions along the \textit{x/y}-axis.  Grid size is 20 cm. b) Log–log plots of the localization error versus the dipole displacement $d$ from the origin indicates that the error roughly scales as $d^3$.  A notable exception to this power law occurs for smaller displacement along the \textit{x}-axis, when the dipole is between the sensor positions - movement away from one sensor position corresponds to movement towards the other position as seen in (a);  in this regime, $d < 30$ cm for \textit{x}-axis data in (b), the localization error is approximately constant.}
\end{figure}
\FloatBarrier
\newpage
\subsection{Validation of a magnetic dipole as the signal source\protect}

There are three different metrics that can be used to validate the source of the measured magnetic field as a dipole, and thereby also substantiate the calculated positions. If the source of the signal resembles a single magnetic dipole, the following model deviation (MD) values should be zero.
\begin{enumerate}
    \item  The direction vectors $\hat{x}$, $\hat{n}_1$, and $\hat{n}_2$ should be coplanar, and their scalar triple product zero,
       \begin{eqnarray} \label{MD1}
       MD_1 = \left( \hat{n}_1 \times \hat{n}_2 \right) \cdot \hat{x}.
        \end{eqnarray} 
    \item For a calculated dipole position, the magnetic fields arising from a dipole, Eq.~\ref{dipole_moment}, can be calculated at the two sensor positions, $\vec{B}_{1c}$ and $\vec{B}_{2c}$, and compared to the actual measured values, $\vec{B_{1}}$ and  $\vec{B_{2}}$.  Comparison of the relative amplitudes, calculated versus measured can be done through a projection, and subtracted from unity, 
    
        \begin{eqnarray} \label{MD2}
        MD_2 = 1-
        \frac{
        \begin{bmatrix} B_1 \\ B_2 \end{bmatrix} \cdot
        \begin{bmatrix} B_{1c}  \\ B_{2c} \end{bmatrix}
        }{
         \sqrt{(B_1^2 + B_2^2)(B_{1c}^2 + B_{2c}^2)}
        }
        \end{eqnarray}
\item Likewise the angular deviation between the measured and calculated fields can be tested by projection.  Since there is a field at each sensor position, the weighted average of the projections are taken and subtracted from unity,
        \begin{eqnarray} \label{MD3}
    MD_3 =  
  1 - 
    \frac{ (\vec{B}_1 \cdot \vec{B}_{1c}) B_1 + (\vec{B}_2 \cdot \vec{B}_{2c})B_2 }{ B_1 + B_2 }.
        \end{eqnarray}
\end{enumerate}
\begin{figure}[t]
\captionsetup{
  format=revjustify,
  singlelinecheck=false
}
\centering
\subfloat[]{\includegraphics[width=0.35\textwidth]{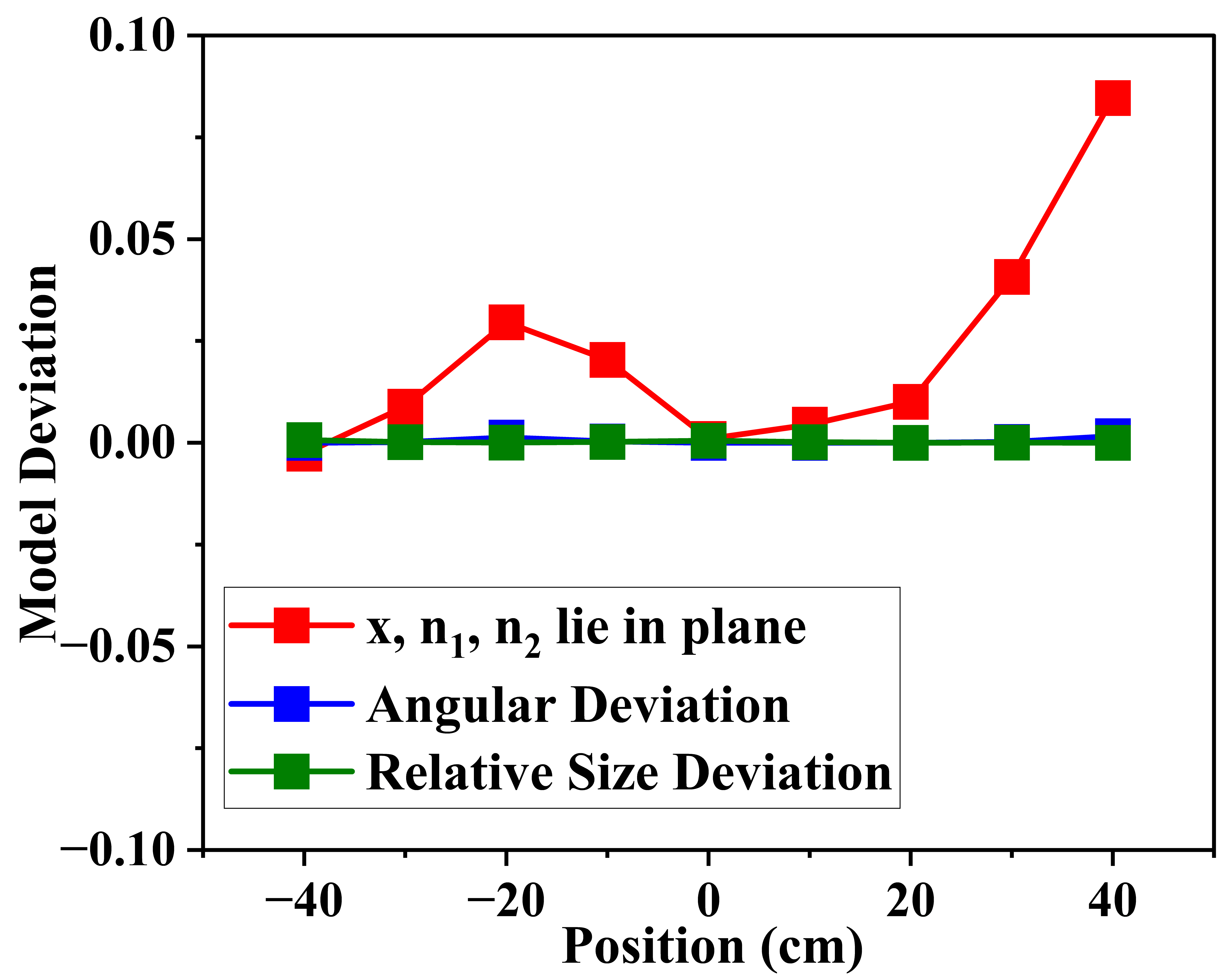}}

\subfloat[]{\includegraphics[width=0.35\textwidth]{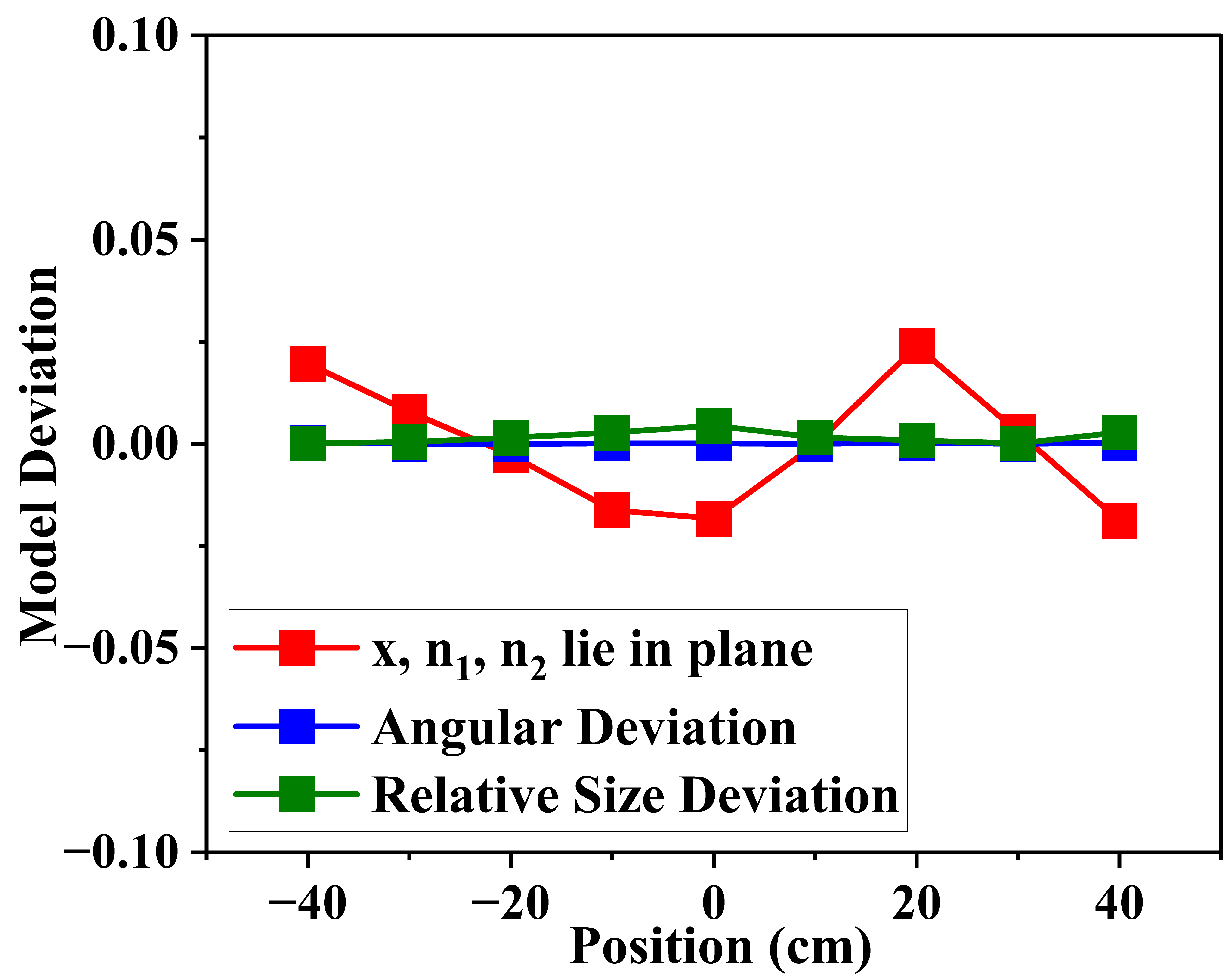}}

\subfloat[]{\includegraphics[width=0.35\textwidth]{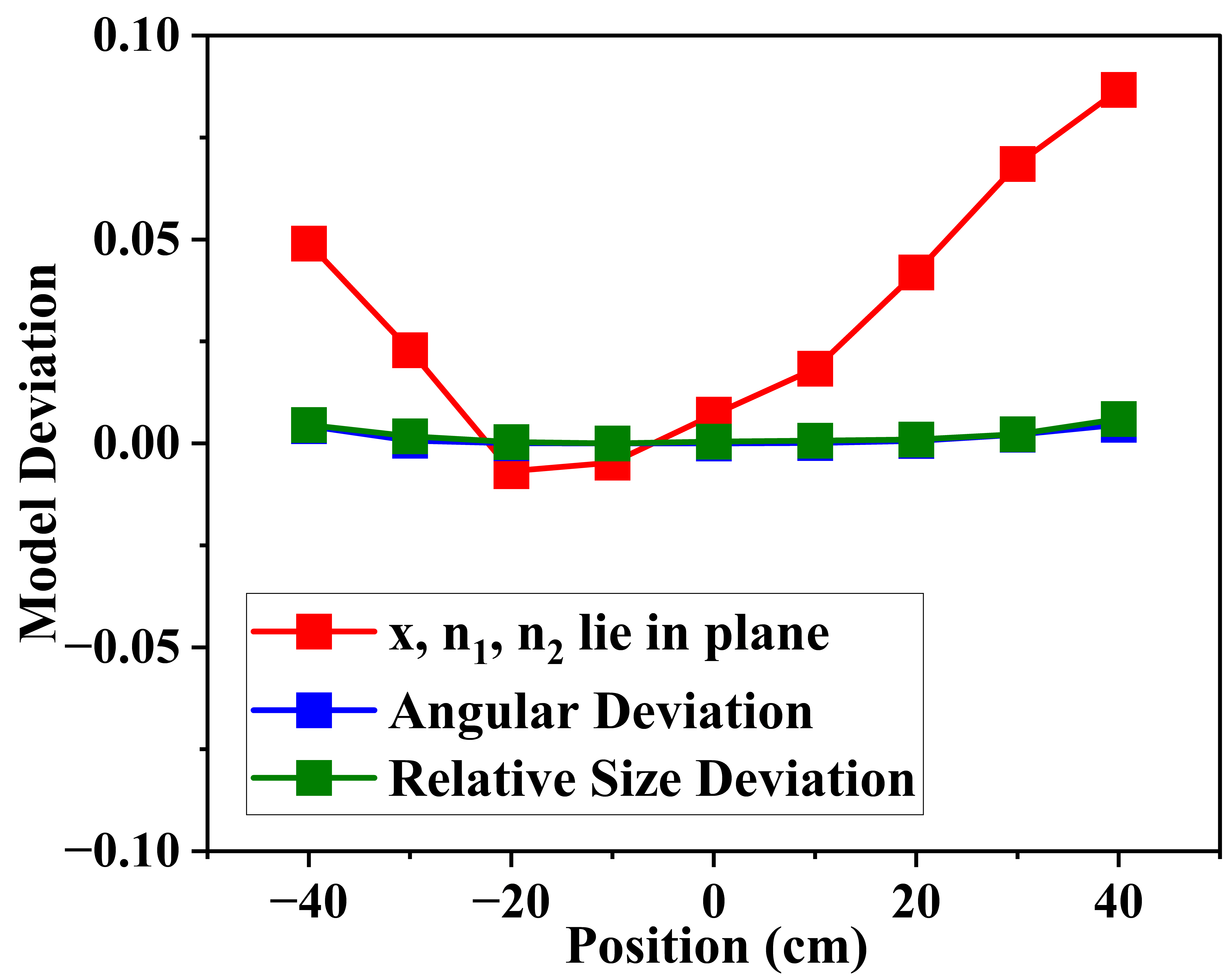}}

\caption{\label{fig:fidelity}Model deviation as a function of displacement along a ruler aligned with \textit{x}-axis, (b) \textit{y}-axis, and (c) and $x/y$ diagonal. In each case, the MD values, given in Eqs.~\ref{MD1}-\ref{MD3}, evaluate the validity of the dipole model.}
\end{figure}

In Fig.~\ref{fig:fidelity}, the three MD values are plotted for the trajectories described in the previous sections.  Values would range from zero to one, with zero indicating data matching a dipole model, and one indicating disagreement.  The values comparing calculated versus measured magnetic fields, $MD_2$ and $MD_3$, are close to zero for all positions, while $MD_1$ shows a small deviation from zero at two of the trajectory edges, a model deviation also observable in Fig.~\ref{fig:Position}.  At the edge is where the fields are weakest and potential interference, most likely self-generated, might create some small corruption of coordinate values. 
 
Overall the experimental results show good agreement with a magnetic dipole model, both through the agreement of physical and calculated dipole coordinates shown in the previous section and the model deviation metrics presented in this section.  Such agreement bodes well for using this methodology to reliably find a local RF source, both for lab and in-the-field applications.\par
The presence of significant interference, which is possible in a noisy unshielded environment, could, however, corrupt dipole localization if no additional data analysis is performed.   Such corruption would be evident in the triplet of model parameters,\textbf{\boldmath${MD = [MD1, MD2, MD3]}$}, deviation away from zero, because of the non-linear and specific nature of the model.    Interference from a far-off source introduces a constant field BI across the two vector measurements positions.   In principle more advanced data analysis could account for this added field by minimizing \textbf{\boldmath$MD(B_1 – B_I, B_2 – B_I)$} as a function of \textbf{\boldmath$B_I$} , thereby recovering the local dipole position from the corrected measurements \textbf{\boldmath$B_1 – B_I$} and \boldmath$B_2 – B_I$.However, implementing such an approach lies beyond the scope of the present work, as it requires simultaneous vector measurements at both positions rather than four sequential measurements with a single sensor. Both simultaneous measurements with a magnetometer array, as well as the mitigation scheme for interference mitigation, highlights a promising avenue for future research.
\FloatBarrier
\section{\label{sec:conclusion}Conclusion\protect }

An oriented magnetic dipole was located in a 80~cm diameter circle in a plane 19.5~cm away from the sensor plane.  The dipole operated at 423~kHz and generated an RF magnetic field of less than a nT at the RF atomic magnetometer. The newly-developed integrated magnetometer, with a sensitivity well below that of ambient magnetic noise, was found to be versatile due to its compactness and flexible wired connections.  This versatility allowed for easy rotation of the sensor for orthogonal measurements.  With two orthogonal measurements, the magnetic field vector was extracted.  Furthermore, an algorithm was developed to determine the coordinates of the dipole from two spatially-separated vector measurements.  The 3D location of the dipole was calculated with error bars from the magnetic field vectors alone and found to be in good agreement with the physical location.  In conclusion, the flexibility and sensitivity of the integrated magnetometer, combined with the demonstrated technique to convert RF magnetic vector measurements to dipole coordinates, is a powerful combination for covert localization of nearby magnetic dipoles with a known orientation.   Potential applications with known orientations include those where a dipole is created according to a pre-determined excitation, for instance NMR or MIT, or an RF tracking tag, restricted to translation or 2D motion.

\section{\label{sec:Acknowledgment}Acknowledgment\protect }
Support for this project comes from the Office of the
Undersecretary of Defense, Director of Defense Research
and Engineering for Modernization (DDRE(M)) Quantum
Science Office Contract 47QFLA23C0002.
\newpage
\bibliography{apssamp}
\end{document}